\documentclass{mn2e}
\usepackage{epsfig}
\begin{document}


\title[Giant comets and mass extinctions]{Giant comets and mass extinctions of life}
\author[W.M. Napier]
{
W.M. Napier$^1$\thanks{E-mail: bill.napier@buckingham.ac.uk(WMN)} \\
$^{1}$Buckingham Centre for Astrobiology, University of Buckingham, Buckingham MK18 1EG, UK \\ 
}

\date{Accepted 2014 December 16. Received 2014 December 12; in original form 2014 March 31}

\pagerange{\pageref{27} -- \pageref{36}} \pubyear{2015}

\maketitle

\label{firstpage}

\begin{abstract}
I find evidence for clustering in age of well-dated impact craters over the last 500 Myr.
At least nine impact episodes are identified, with durations whose upper limits are set by
the dating accuracy of the craters. Their amplitudes and frequency are inconsistent with an
origin in asteroid breakups or Oort cloud disturbances, but are consistent with the arrival and
disintegration in near-Earth orbits of rare, giant comets, mainly in transit from the Centaur
population into the Jupiter family and Encke regions. About 1 in 10 Centaurs in Chiron-like
orbits enter Earth-crossing epochs, usually repeatedly, each such epoch being generally of a
few thousand years’ duration. On time-scales of geological interest, debris from their breakup
may increase the mass of the near-Earth interplanetary environment by two or three orders
of magnitude, yielding repeated episodes of bombardment and stratospheric dusting. I find
a strong correlation between these bombardment episodes and major biostratigraphic and
geological boundaries, and propose that episodes of extinction are most effectively driven
by prolonged encounters with meteoroid streams during bombardment episodes. Possible
mechanisms are discussed.
\end{abstract}


\section {Introduction}

The discovery of an excess of iridium at the Cretaceous-Tertiary boundary of 65 Myr BP, and its interpretation as the result of an asteroid impact (Smit \& Hertogen 1980, Alvarez et al 1980), lent support to the hypothesis that large-body impacts are the prime cause of mass extinctions on Earth (Napier \& Clube 1979, Raup 1991, Rampino 1994). However subsequent studies have not, with the still-debated exception of the Cretaceous-Tertiary event itself, confirmed this hypothesis. For example the Chesapeake (40~km) and Popigai (90~km) craters are amongst the largest known terrestrial impact structures, dated to $\sim$35.5~Myr, in the late Eocene, but are associated with only minor, regional extinctions of a few radiolarian species. The 80~km Manicougan crater, aged 214~Myr, is likewise associated with only minor extinctions. Matsumoto \& Kubotani (1996) found that the mass extinction and impact crater records are correlated with only 93-97\% significance. 

Considerable attention has been paid to the question of periodicity in the impact cratering record, and whether, if such exists, it can be connected with Galactic cycles or mass extinction episodes; the issue remains controversial (cf Bailer-Jones 2009, 2011, Melott \& Bambach 2013). Less attention has been paid to the issue of \textit{episodes} of bombardment, although several such have been noted and claims have been made of correlation with mass extinction (Keller et al 2012; McGhee 1996).  Comet showers induced by stellar penetration of the inner Oort cloud, or asteroid showers deriving from the breakup of main belt asteroids, are inadequate by about an order of magnitude to reproduce the sharpness, amplitude and frequency of the bombardment episodes (Napier \& Asher 2009).  Leakage from the main asteroid belt appears in any case to be inadequate by an order of magnitude to supply near-Earth object impactors say more than 3 or 4 km across (Menichella et al 1996, Minton \& Malhotra 2010). It has been proposed that bombardment episodes are caused by the disintegration of giant comets in short-period, Earth-crossing orbits (Clube \& Napier 1984). A more recent worldwide extinction of species at 12,800 BP, accompanied by a climate cooling which lasted for over 1000 years, has likewise been attributed to an encounter with the debris from such a body (Napier 2010; Wittke et al 2013; Moore et al 2014), although this remains a controversial proposition (van Hoesel et al 2014). Models that neglect the injection of large comets into the near-Earth environment (e.g Chapman 2004, Boslough et al 2013) fail to yield large-scale catastrophic disruption of this sort in the recent past, but it remains to be seen how often such comets arrive and what effect they may have on the biosphere over geological as well as human timescales; this is the subject of the present paper.

\section{Transfer into short-period orbits}

Recent years have seen the discovery of large populations of apparently icy bodies, some in unstable orbits, at the edge of the planetary system. Hahn \& Bailey (1990) have demonstrated that 95P/Chiron, the first Centaur to be discovered, is in a chaotic orbit and may have entered a short-period, Earth-crossing configuration several times over the past $\sim$10$^5$~yr. From Spitzer observations its diameter is estimated to be  233$\pm$14~ km (Stansberry et al 2008).  The potential for icy bodies of 100 km diameter upwards to enter short period orbits with perihelia $q\le$1\,au has been demonstrated both theoretically (Hahn \& Bailey loc. cit.) and  observationally. Sekanina and colleagues (2000, 2003, 2004, 2007) have shown that the Kreutz group of sungrazing comets is derived from a complex hierarchy of fragmentations, starting with the disintegration of a progenitor comet of probable diameter $\sim$100~km about 1700 years ago. Likewise an extensive study of the Taurid complex supports the view that it originated from the breakup of an exceptionally large comet in the distant past (Whipple 1940, Sekanina 1972, Clube \& Napier 1984, Asher et al 1994). The currently observed system may be at least 20,000 years old (Steel et al 1991), but an injection in a much more distant past is also possible.

A 100~km comet striking the Earth would carry $\sim$1000 times the energy involved in the creation of the 150~km Chicxulub crater and would presumably remove the surface biosphere; impact crater statistics show this to have been an unlikely event over the 600~Myr timescale of the Phanerozoic. However, disintegration seems to be a normal if not predominant mode of comet loss. Given that a Chiron-sized body has a mass 10$^{21}$-10$^{22}$~gm, as against $\la 10^{19}$~gm for the entire current near-Earth asteroid system, there is the potential for a substantially enhanced hazard during the breakup of such a body in the inner planetary system. Such a breakup might involve multiple bombardment episodes and substantial dust input (Clube \& Napier 1984), leading perhaps to a qualitatively different astronomical scenario for mass extinctions from that of a single, 10~km asteroid impact (Alvarez et al 1980).

\subsection{Numerical integrations}

Di Sisto \& Brunini (2007) modelled the Centaur population on the assumption that it derives from the scattered disc one, following the evolution of 1000 objects for 4.5 Gyr and estimating the population by scaling from an assumed scattered disc population of 7.5$\times 10^9$ bodies with radii $R>$1~km (Fern\'{a}ndez et al 2004). They deduced a population of 2.8$\times 10^8$ Centaurs, with an order of magnitude uncertainty. The mean dynamical lifetime $L_d$ of Centaurs was found to decline strongly with perihelion distance, with $L_d<$10~Myr for objects with $q\le$17\,au, down to 7.6~Myr for bodies with 5.2$<a<$30\,au and $\sim$6.2\,Myr for Centaurs defined as $Q<$40\,au and $a<$29\,au. Horner et al (2004a, 2004b) obtained $L_d\sim$2.7~Myr for a sample of Centaurs defined by the latter criteria. 

The hazard represented by the arrival of large comets is determined in part by their rate of arrival in the near-Earth environment rather than their mean lifetime as Centaurs, and this rate is dominated by that of the most dynamically unstable bodies at any given time. Horner et al (2004b), simulating the evolution of Chiron with 729 clones for up to 3~Myr, found that over this time, more than half the clones became short-period comets at some point, and 85 became Earth-crossers. Similar behaviour occurred in other Centaur clones with similar orbits.

\begin{figure}
\includegraphics[angle=270,width=\linewidth]{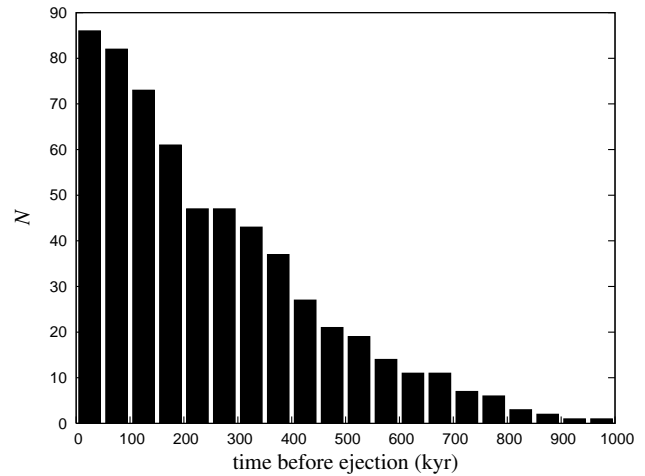}
\caption{Survival times of 86 out of 100 Chiron clones before ejection from the solar system.}
\label{fig:ejection}
\end{figure}

\begin{figure}
\includegraphics[angle=270,width=\linewidth]{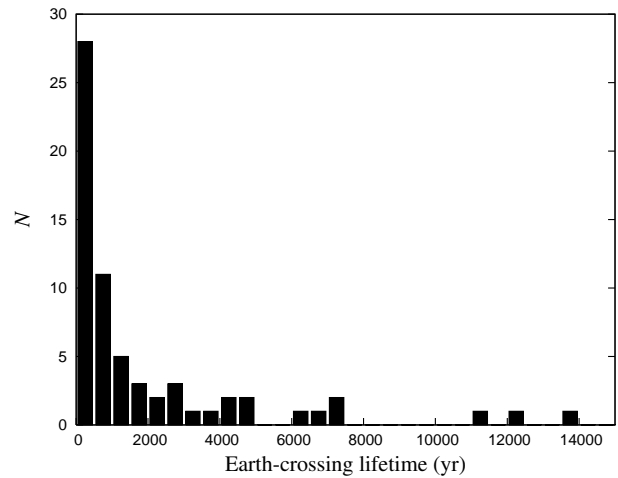}
\caption{ Distribution of the time spent by 100 Chiron clones in orbits with $q <1$ au. The 65 Earth-crossing epochs were due to 9 clones entering and re-entering these orbits.}
\label{fig:earthlife}
\end{figure}

\begin{figure}
\noindent
\begin{minipage}[b]{\linewidth}
\includegraphics[angle=270,width=\linewidth]{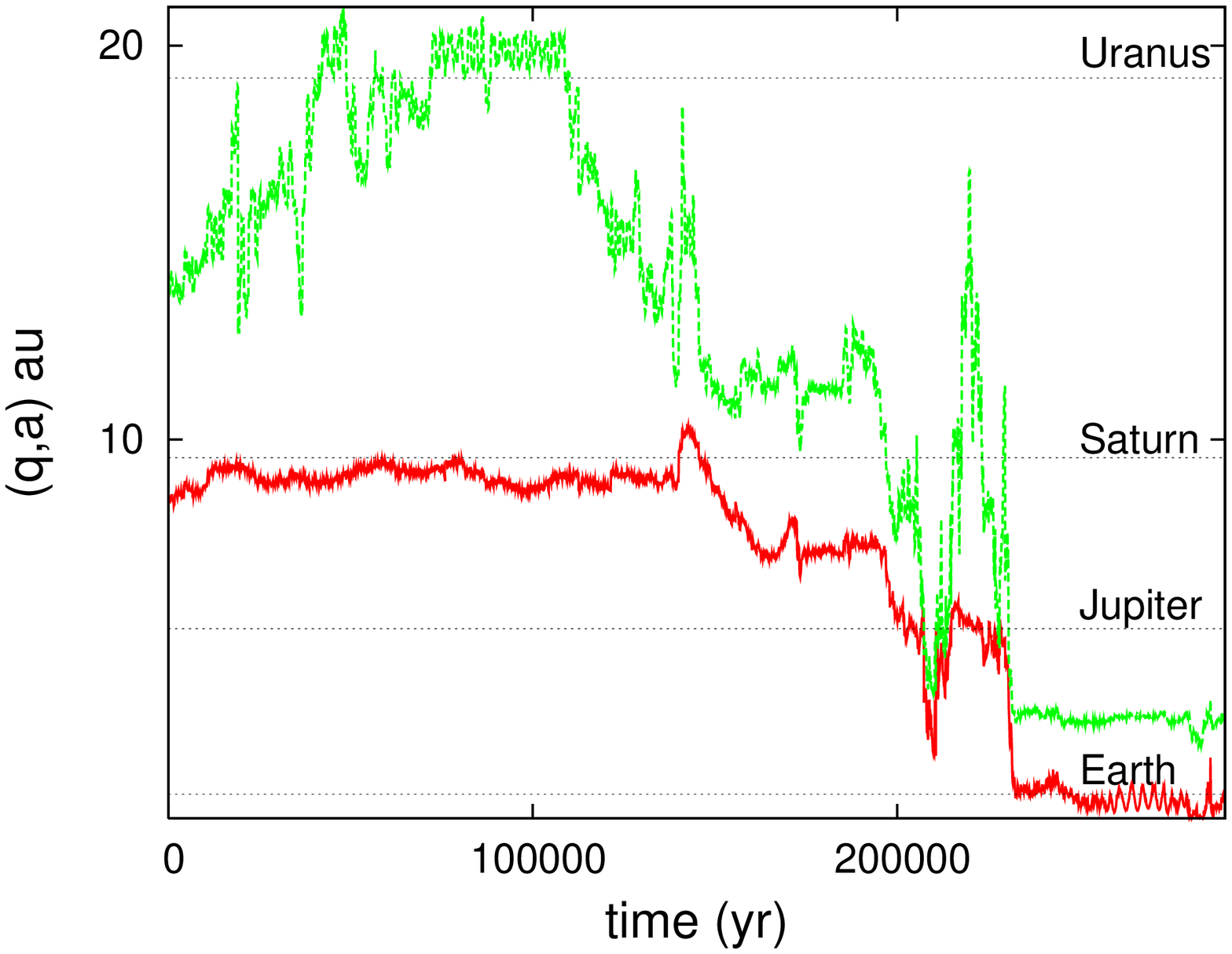}
\end{minipage}
\noindent
\begin{minipage}[b]{\linewidth}
\includegraphics[angle=270,width=\linewidth]{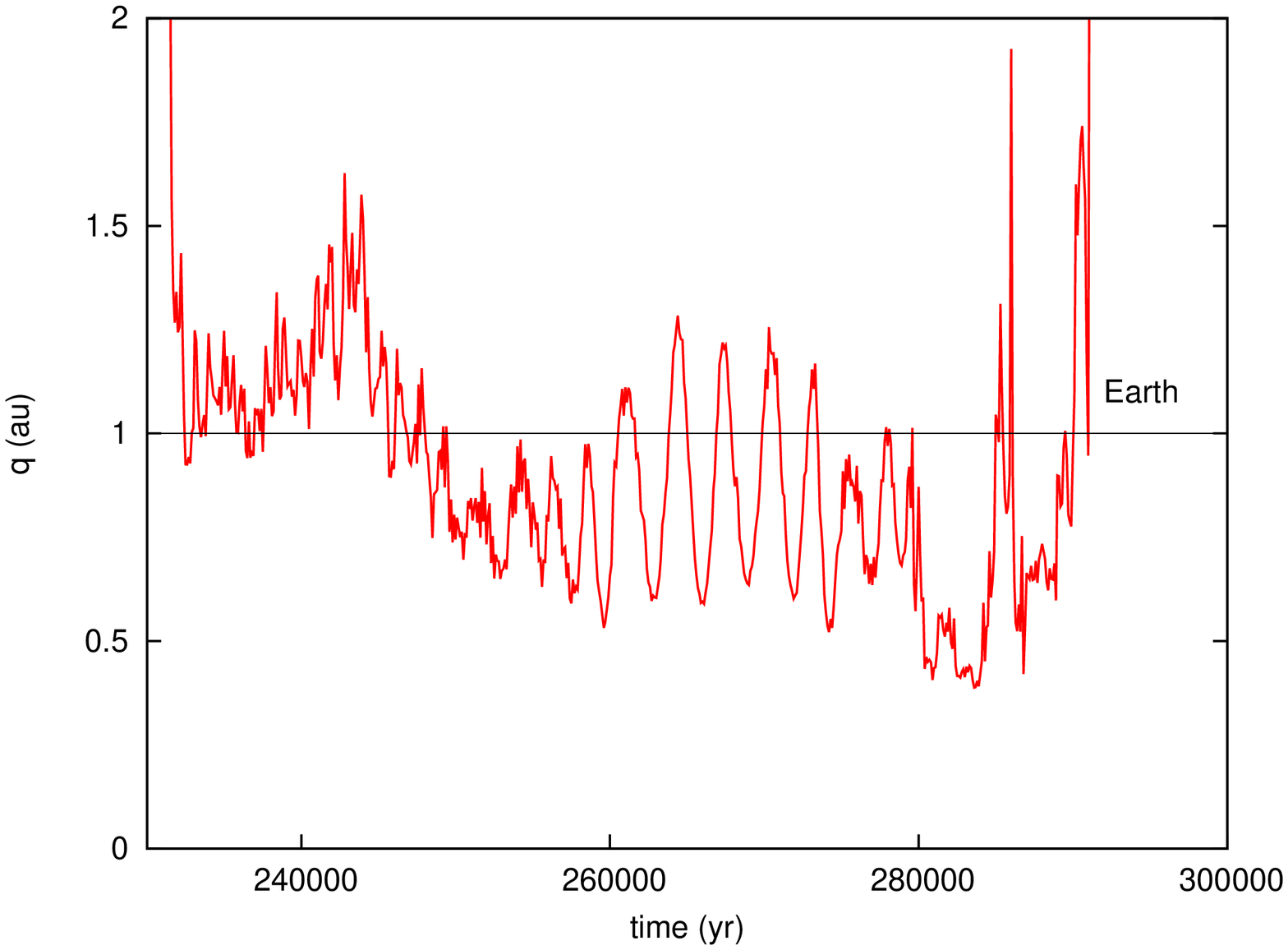}
\end{minipage}
\caption{Top: Chiron clone to Earth-crosser. The upper curve represents the semimajor axis, the lower one is the perihelion. Bottom: The last 60,000~yr in more detail; substantial disintegration is likely to take place in this orbital state.}
\label{fig:chiron}
\end{figure}

I carried out orbital integrations of Chiron clones over 1~Myr with the Mercury software developed by Chambers  (1999), using the same input conditions as Horner et al (2004b). All the planets were incorporated. The hybrid part of the package was used (employing a symplectic integrator, but switching to Bulirsch-Stoer for close encounters). The accuracy parameter was set to 10$^{-10}$ and initial timestep was 60 days; over a typical run this conserved energy to a fraction ${\sim} 1.7 \times 10^{-6}$ and angular momentum to $10^{-12}$. Of 100 clones, 86 were ejected within 1~Myr, their mean dynamical lifetimes before ejection being 0.35 Myr (Fig.~\ref{fig:ejection}). About half temporarily entered short-period orbits, and nine temporarily entered Earth-crossing orbits, often repeatedly. There were 65 Earth-crossing epochs in all, ranging from $\sim$100 to $\sim$14000~yr, with mean duration $\sim$2000 yr (Fig.~\ref{fig:earthlife}). The dynamical evolution of one such clone, which included a short-period phase ($P\sim$3.5-5.3~yr) of 50,000 year duration, is illustrated in Fig.~\ref{fig:chiron}. In these runs the bodies are treated as point masses and their physical evolution is neglected. If 100 simulated Chirons, with mean lifetime 0.35~Myr, yield 65 Earth-crossing episodes, then the  frequency of such episodes is 65/100/0.35 Chiron$^{-1} \rm Myr^{-1}$, giving recurrence time $\sim$1.9 Myr per Chiron clone.

\subsection{Estimates of the Centaur population}

Di Sisto \& Brunini (2007) estimated the number of Centaurs larger than Chiron to be 360$\le N\le$650, of which there are 4$\le N\le$7 within 18\,au of the Sun (Centaurs are here defined as objects with perihelia between Jupiter and Neptune, irrespective of semi-major axis). This is consistent with the discovery of four Centaurs larger than Chiron, three of them within 18\,au of the Sun. If there are typically four objects of this size at any given time with similar orbital properties, each with arrival probability into an Earth-crossing orbit of $p\sim$0.5~Myr$^{-1}$, then there is an expectation of one Chiron-sized Earth-crosser every half a million years or so. 
 
The size distribution of Centaurs may depend on the relative contributions from their source populations. For long-period comets, Fern\'{a}ndez \& Sosa (2012) find a cumulative radius $R$ distribution $N(R)\propto R^{-s}$, with $s = 2.15\pm 0.75$, consistent with $s=2$ for an evolved population in the diameter range 20-200 km, expected from current models of planetesimal growth (e.g. Kenyon \& Bromley 2012). Bauer et al (2013) analysed thermal infrared observations of 52 centaurs obtained from the Wide-field Infrared Survey Explorer and obtained a biased power law size distribution with index $s=2.7\pm0.3$. If we adopt a diameter $D$=240\,km (Stansberry et al 2007), then with these distributions the number of comets in Chiron-like orbits, $\ge$100\,km diameter, entering Earth-crossing orbits in a million years is $\sim$6-14 times the Chiron rate of entry, or about one every 36,000-86,000 years. This figure is likely to be conservative as additional contributions are expected from more remote Centaurs. Asher \& Steel (1993), for example, followed the evolution of the large Centaur 5145 Pholus over 1~Myr using 27 clones, which yielded two particles entering Earth-crossing orbits for several tens of thousands of years, implying that a body in such an orbit has a 5-10\% chance becoming Earth crossing within a million years. 

Defining the system of Jupiter family comets (JFCs) as having orbital periods $P<$20~yr and Tisserand criterion $T<$3, Di Sisto et al (2009) estimate a population of $N$=450 with perihelia $q\le$2.5~au. For a mass distribution $n(m)\,dm \propto m^{-\alpha}\,dm$ in the radius range 1$<R<$10~km, the mass of the JFC system is then
\begin{equation}
M_{JFC} = N(\frac{\alpha-1}{2-\alpha})\frac{m_{10}}{m{_1}}m_1
\label{eq:JFCmass}
\end{equation}
where ($m_{10},m_1$) are respectively the masses corresponding to JFCs of radius (10~km, 1~km) respectively. The size distribution adopted by Di Sisto et al translates to a population index $\alpha=1.9$. With this, and $m_{10}/m{_1} = 1000$, we find $M_{JFC}\sim$8100$m_1$. A comet of radius 1~km and density $\rho = 0.5$~gm\,cm$^{-3}$ has mass $m_1\sim$2.6$\times 10^{14}$~gm. Hence $M_{JFC}\sim$2.2$\times 10^{18}$~gm. A similar exercise for the current near-Earth asteroids (NEA) yields, if they are largely composed of rock with density 2.5~gm\,cm$^{-3}$, a mass $\sim$5$\times 10^{18}$~gm. A 250~km comet with density 0.5~gm\,cm$^{-3}$ has mass $\sim$5$\times 10^{21}$~gm, two thousand times that of the current JFCs and a thousand times that of the NEAs. Thus in terms of Earth-crossing mass the present-day impact hazard represents, not a steady state, but a snapshot of a strongly flickering system. The fluctuation timescale is $\sim$0.5 Myr for bodies of at least Chiron size, and $\sim$0.03-0.1~Myr for comets at least 100~km in diameter. 
\section{Breakup modes}

\begin{figure}
\includegraphics[angle=270,width=\linewidth]{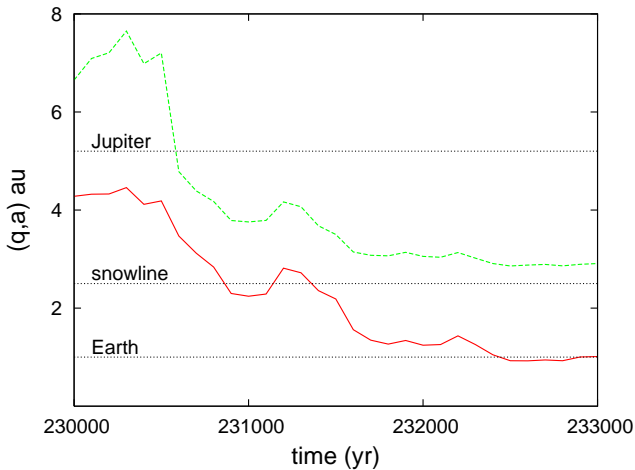}
\caption{Part of Fig.~\ref{fig:chiron} at higher resolution still, showing a transition from water snowline (sublimation at $q\la$2.5\,au) to a short-period orbit with $q\le$1\,au in less than 1000 years.}
\label{fig:snowline}
\end{figure}

 The outgassing of an active Chiron-sized comet in say a 5~yr orbital period (Fig.~\ref{fig:chiron}) is equivalent to that of 15,000 Halley-sized comets in Encke-like orbits, or in excess of 40 strong meteor showers per night, and the same by day. Apart from periods of dormancy, this will persist over the dynamical or physical residence time of the comet, whichever is shorter. Such intense atmospheric bombardment is expected to have consequences for the biosphere (Section~\ref{sec:terr}).  This decay mode may be checked, albeit temporarily, by the development of a refractory porous crust.  In the rubble mantle model of Jewitt (2002), the timescale for mantle growth for a water ice and carbon monoxide composition is of order a year at 2.5\,au and $\sim$100~yr at 5\,au.  The transfer of a Centaur from an orbit with $q = 2.5$\,au (the water snowline) to an Earth-crossing orbit may take place in less than 1000 years (Fig.~\ref{fig:snowline}).  

`Spontaneous' splitting or fragmentation of very large comets are probably unimportant disintegration routes since their escape velocities ($\sim$70 m\,s$^{-1}$ for a 250 km comet, $\sim$30 m\,s$^{-1}$ for a 100 km one) are in excess of the likely separation speeds of fragments. Disintegration is more likely to take place through sublimation, or tidal disruption by the Sun or Jupiter. The simplest breakup mode, namely generation of an intense meteoroid stream along the track of the comet, has been discussed by Clube \& Napier (1984).  According to \"{O}pik (1951), a comet loses $4/\sqrt{q}$~m of surface per revolution, a figure not very different from that found empirically by Di Sisto et al (2009). This yields an output of 10$^{18}$~gm of gas and dust per revolution for a Chiron-sized body in an Encke-like orbit, but could rapidly produce a dormant stage, so preserving the integrity of the comet as its perihelion is driven downwards (e.g. Rickman et al 1990). 

Tidal disruption by Jupiter is also a common process, several examples having occurred in recent times (e.g. D/1993 F2 Shoemaker-Levy 9). The timescale for collision of a 1~km JFC on Jupiter is about a decade (Zahnle et al. 1998, Sanchez-Lavega 2010), implying that passage of a comet within its Roche limit (${\sim}3.4R_J$) is virtually an annual event. It is thus likely that the JFC system is replenished sporadically by the disintegration of large comets passing within the Jovian Roche lobe  (Bredichin 1889). Pittich \& Rickman (1994) have shown that fragments of a JFC so disrupted would lose their memory of a common origin within a few thousand years. A $\sim$10$^3$  enhancement of the JFC population by the disruption of a 250~km comet would feed into a corresponding increase in the NEO cometary population over the physico-dynamical decay time of the JFC.  

A common fate of comets thrown into short-period orbits, if not disintegration through sublimation, is tidal disruption by the Sun (e.g. the $\sim$100~km Kreutz progenitor). Destruction through tides is a volume rather than a surface phenomenon, and is only weakly dependent on the size of the comet. Bailey et al (1992) have shown that, under the influence of slowly acting secular perturbations of the giant planets, $\sim$5-15\%  of comets with initial perihelion distances $q\le$2~au may evolve into a sungrazing or sun-colliding state. This evolution occurs at intervals of order 10$^3$ revolutions, about a tenth of the dynamical ejection timescale.  The tidal disruption of a comet entering the Roche limit, ${\sim}3.5 R_\odot$, may create a hierarchy of fragmentations. Passage through such clusters, before they disperse into the background, will generate one or more series of sharp bombardments, analogous to meteor storms but of much greater intensity, throughout the lifetime of the comet.

\subsection{Encke-like orbits}

Ipatov \& Mather (2004), following the orbital evolution of 26,000 fictitious Jupiter-crossing objects, found that bodies in orbits close to that of Comet 2P/Encke had mean collision probabilities with the terrestrial planets about two orders of magnitude greater than those of other comets, remaining in typical NEO orbits for millions of years. Levison et al (2006), following orbits from the Kuiper belt, deduced that at any one time there should be one to a dozen objects in Encke-like orbits, neglecting both physical decay and nongravitational effects. They found that the time taken for a JFC to attain an Encke-like orbit, dynamically uncoupled from Jupiter, is 0.1~Myr, two orders of magnitude longer than its physical lifetime, but proposed that 2P/Encke might have survived destruction through a period of dormancy. 

Unless dormancy is 100\%, however, non-gravitational forces may operate and enhance the rate of transfer substantially in combination with mean motion resonances with Jupiter (Steel \& Asher 1996; Fernandez et al 2004; Pittich et al 2004). The acceleration due to such forces varies as area/mass, i.e. $R^2/R^3\propto1/R$, but the fuel available $\propto R^3$ and unless permanent dormancy intervenes, long term non-gravitational perturbations (acceleration times time) are thus more effective in perturbing large comets ($\propto R^2$). In the event of a perihelion disintegration, for example, the cross-sectional area of fragments reaching the Earth's distance is typically about 10,000 times that of the comet itself, for an initial dispersion speed of 5 m\,s$^{-1}$ (Napier 2010). However the transition probability from the JF to Encke domains remains uncertain to order of magnitude. 

In summary, on the arrival of a giant comet in an Earth-crossing orbit, whether Jupiter family or Encke-like, several modes of disintegration are possible which may act to enhance the terrestrial bombardment rate by two or three powers of ten over background, during its physico-dynamical lifetime. Direct impact of an incoming large comet is much less probable than interaction with its debris, which may include many short-lived fragments, and a correspondingly spiky distribution of terrestrial impactors is expected. This can be tested by examining the impact cratering record. 

\section{Bombardment episodes}

\begin{table}
\centering
\begin{tabular}{lllll}  \hline
$D$ (km) & age (Myr)       & crater              &   no. & epoch (Myr)\\ \hline
30--120 & 0.78$\pm$0.1   & Austr. tektites &  1     & \\  
14	    & 0.9$\pm$0.1         & Zhamanshin  	&  1     & \\
10.5 	& 1.07$\pm$0.0       &  Bosumtwi     	& 1      & \\
 3.4	  & 1.4$\pm$0.1         & New Quebec  & 1      &  1.1$\pm$0.03 \\  [4pt]
	
 52	    & 2.5$\pm$2.5         & Kara-Kul          &       	& \\
 18	    & 3.5$\pm$0.5         & El'gygytgyn  	&         & \\
 10	    & 5$\pm$1               & Karla            	&         & \\
  8	  	& 5$\pm$3               & Bigach          	&         & \\
 24     & 15.1$\pm$0.1       &  Ries               &         & \\ 		               [4pt]

 40     & 35.3$\pm$0.1       &  Chesapeake  & 2      & \\ 
 90     & 35.7$\pm$0.2    	 	& Popigai          & 2      & \\
 28     & 36.4$\pm$4      	 	&  Mistastin      &  2   	& \\
  7.5   & 37.2$\pm$1.2  	   	& Wanapitei   	&	2      & 35.4$\pm$0.2 \\  [4pt]
	
 15     & 42.3$\pm$1.1  	  	& Logoisk          &	        & \\    [4pt]

  9      & 46$\pm$3         	  & Ragozinka     & 3	     & \\
 25     & 49.0$\pm$0.2       & Kamensk       &  3     & \\        	
 45     & 50.50$\pm$0.76   & Montagnais   & 3      &  49.1$\pm$0.2 \\      [4pt]

 12.7  & 58$\pm$2             &  Marquez  	    &         & \\  [4pt]

150    & 64.98$\pm$0.05		& Chicxulub      & 4      & \\     
 24     & 65.17$\pm$0.64		& Boltysh          &  4	   & 65.0$\pm$0.05  \\  [4pt]

 65     &70.3$\pm$2.2 				& Kara              &         & \\
 35     & 74.1$\pm$0.1 			& Manson        &         &   \\  
  6.5   & 81.0$\pm$1.5 			&	 Wetumpka  &         & \\	
	19     & 89.0$\pm$2.7 			& Dellen           &         & \\
  13     & 99$\pm$4  					&  Deep Bay    &        	& \\  [4pt]

  9      & 121.0$\pm$2.3			& Mien                  &  5  & \\
 12     &123$\pm$1.4  				&	Vargeao Dome  &  5 & 122.5$\pm$1.2 \\  [4pt]

 55     &  128$\pm$5   				& Tookoonoka       & 	   & \\  [4pt]

 40     & 142.0$\pm$2.6 		&	 Mjolnir             & 	6  & \\
 22     & 142.5$\pm$0.8 		&	 Gosses Bluff     &    6  & \\  	
 70     & 145.0$\pm$0.8 		& Morokweng     	&    6  & 143.7$\pm$0.6 \\  [4pt]

  3.2   & 165$\pm$5   				& Zapadnay            &  7   & \\
 40     & 167$\pm$3   				& Puchezh-Kat.       &  7   & 166.5$\pm$2.6 \\  [4pt]

40--50 & 200$\pm$1          &  seismite             & 8   & \\
 23     & 201$\pm$2  					&  Rochechouart    & 8    & 200.2$\pm$0.9 \\  [4pt]

85      & 214$\pm$1 					&  Manicouagan     &       & \\ 	
 40     & 254.7$\pm$1.3      &  Araguainha         &       & \\ [4pt]
 
 52     & 376.8$\pm$1.7  		& Siljan                   & 	  9   & \\
  8.5   & 378$\pm$5  					&  Ilyinets                &   9  & \\ 
 15     & 380$\pm$5  					&  Kaluga                & 	9  & 377.2$\pm$1.5 \\  [4pt]

  6      & 445$\pm$2   				& Pilot   &  	& \\  \hline

\end{tabular}
\caption{Impact structures, selected as discussed in the text. Data from the Earth Impact database, with the addition of the Australasian tektites and a seismite impact marker at 200~Myr. Possible impact episodes  (no.) are listed, epochs being given as the weighted mean age of the associated craters, with an unbiased weighted estimate of the sample standard error. Updated from Napier (2006); cf Napier (2014). }
\label{ta:craterlist}
\end{table}

\begin{figure}
\includegraphics[angle=270,width=\linewidth]{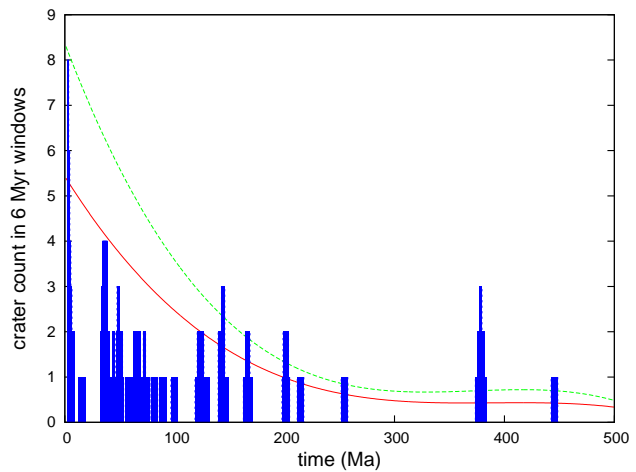}
\caption{Age distribution of Table~\ref{ta:craterlist} impact structures. Data have been smoothed by a rectangular window 6~Myr wide, in 1~Myr steps. Estimates of the relative rate of decline as one goes into the past are shown by the curves, derived as described in the text; the upper curve is the Sturges fit, the lower is the Rice one.}
\label{fig:craters}
\end{figure}

Of the 184 confirmed craters in the Earth Impact Database, accessed in 2014 September, 39 were culled with $D>$3~km, dating accuracy $2\sigma\leq$5~Myr, age $\leq$500~Myr. Two non-crater impact structures were added: (a) Simms (2003, 2007) has identified a seismite layer which appears to have been created from another large impact, yielding perhaps a 40-50 km unidentified crater, at least 600 km west or northwest of central Britain; and (b) the Australasian tektites have been added; their source crater too has not been identified, but published estimates of its size range from 30 to 120 km (Lee \& Wei 2000).  Thus the database comprises 41 well-dated impact structures covering the last 500 Myr. This is listed in Table~\ref{ta:craterlist} and shown in Fig.~\ref{fig:craters}. The solid lines are estimates of the decline of observed crater numbers with time, and are cubic fits to histograms of the database with bin widths respectively $k$ = 70~Myr and 100~Myr. There is no unique optimum histogram giving a broad trend; the former is optimal under the `Rice rule'  $k = 2 n^{1/3}$, $n$ the number of data (Lane 2014), and the latter derives from the rule of Sturges (1926), namely $k = \log_2 n +1$.  A somewhat larger dataset ($n =50$ craters with $\sigma\leq $10 Myr)) was taken in obtaining these secular trends. Discovery completeness declines to 50\% within 120~Myr for impact craters with diameters $D>$40~km, and within 70~Myr for smaller craters. Identification of impact craters with $D>$20~km is only $\sim$15\% complete over the 500 million year period, declining to $\sim$5\% complete for impact craters $>$10 km across. Nearly all such craters have been discovered on land. Evidently the record of impact craters is erased by erosion, sedimentation and subduction. In Table~\ref{ta:craterlist} we define the epoch of an episode by the weighted average date of the impact craters belonging to it. The intrinsic duration of each bombardment episode could be as little as a few millennia or as much as 1 or 2~Myr if caused by the breakup of a single, large Jupiter family or Encke-type comet, for example via the Jovian Roche lobe.

\begin{figure}
\noindent
\begin{minipage}[b]{\linewidth}
\includegraphics[angle=0,width=\linewidth]{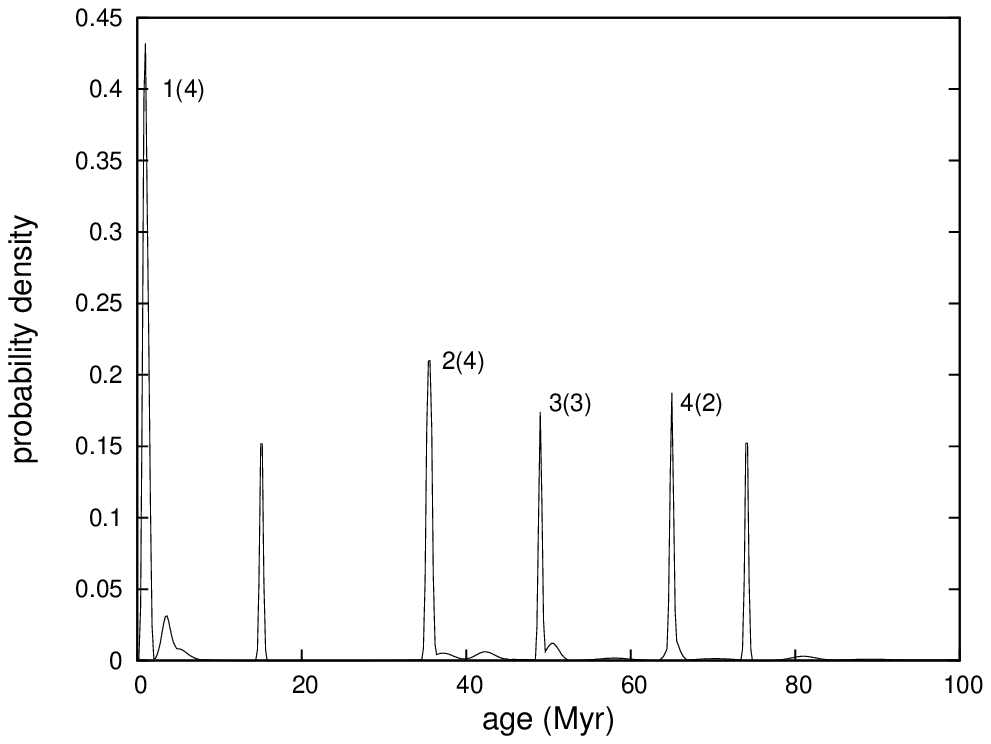}

\includegraphics[angle=0,width=\linewidth]{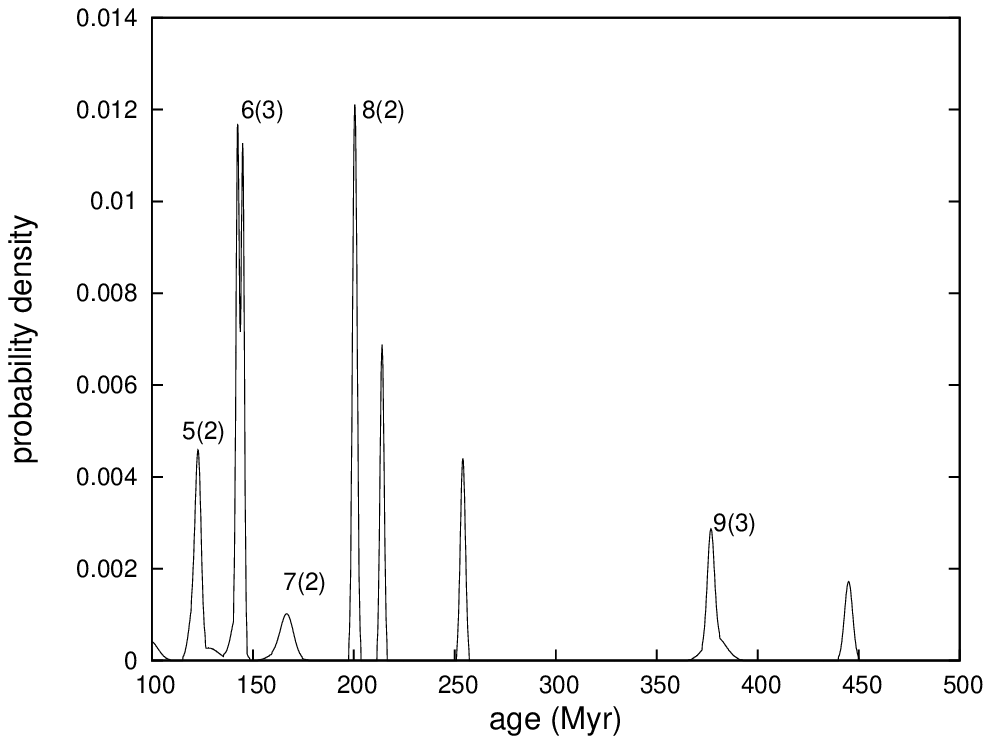}
\end{minipage}
\caption{Probability density distribution of impact structures listed in Table\ref{ta:craterlist}, using Gaussian kernel estimators corresponding to individual standard errors. Top: 0-100 Ma; bottom: 100-500 Ma.  The putative multiple impact episodes of Table\ref{ta:craterlist} are labelled, with the corresponding number of impact structures in brackets.}
\label{fig:kernel}
\end{figure}

\begin{figure}
\noindent
\begin{minipage}[]{\linewidth}
\includegraphics[angle=0,width=\linewidth]{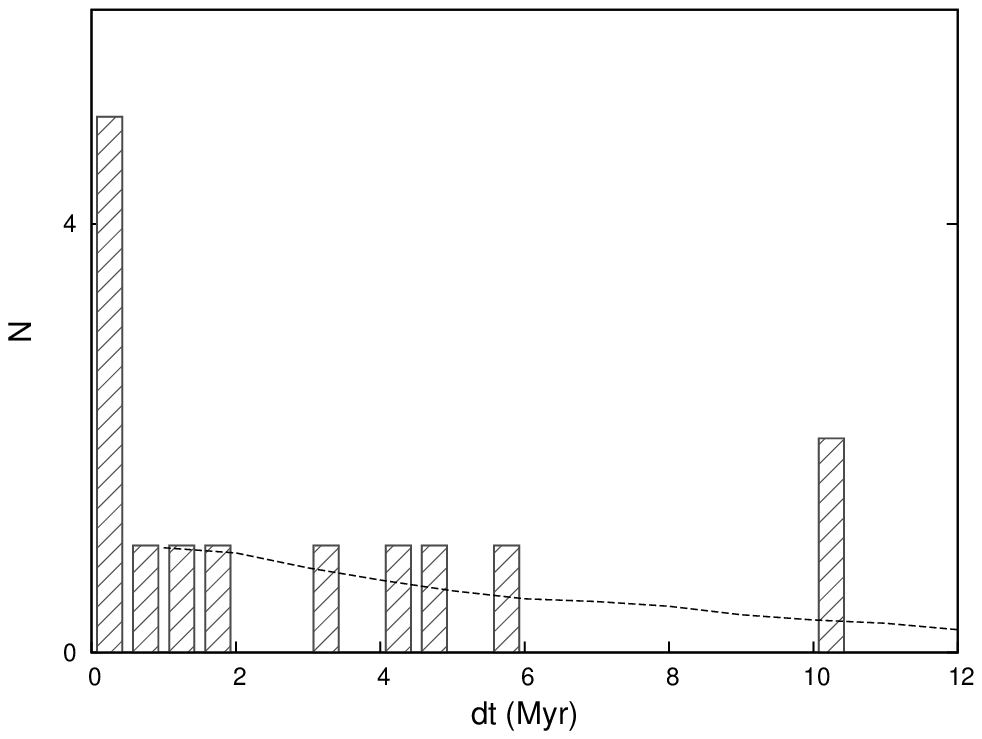}
\end{minipage}
\noindent
\begin{minipage}[b]{\linewidth}
\includegraphics[angle=0,width=\linewidth]{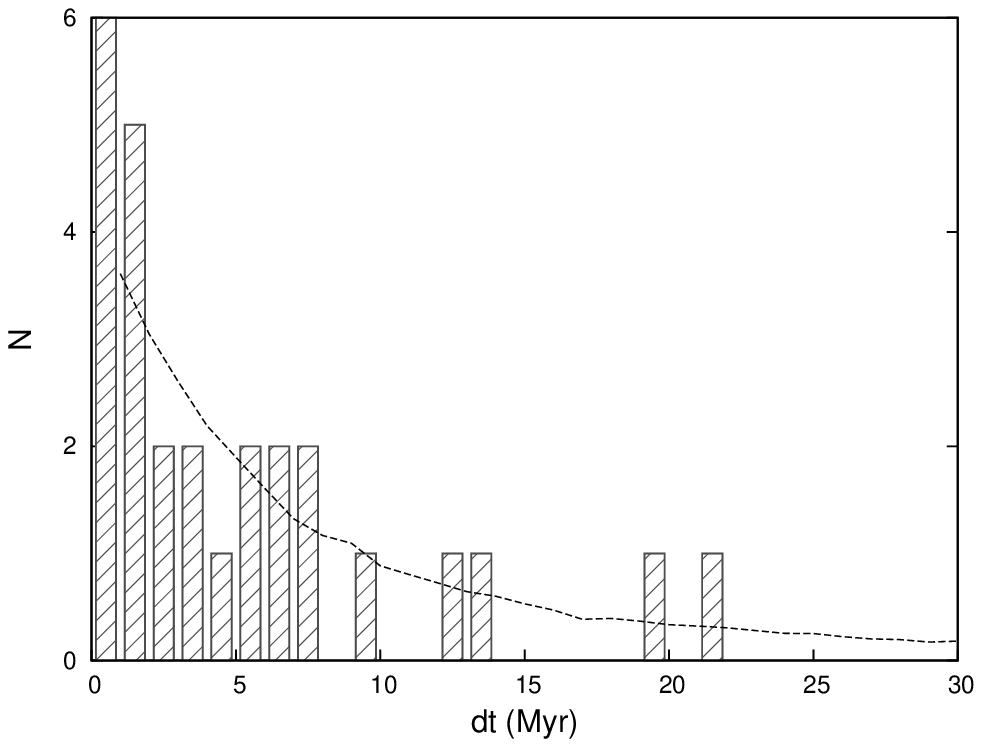}
\end{minipage}
\noindent
\begin{minipage}[b]{\linewidth}
\includegraphics[angle=0,width=\linewidth]{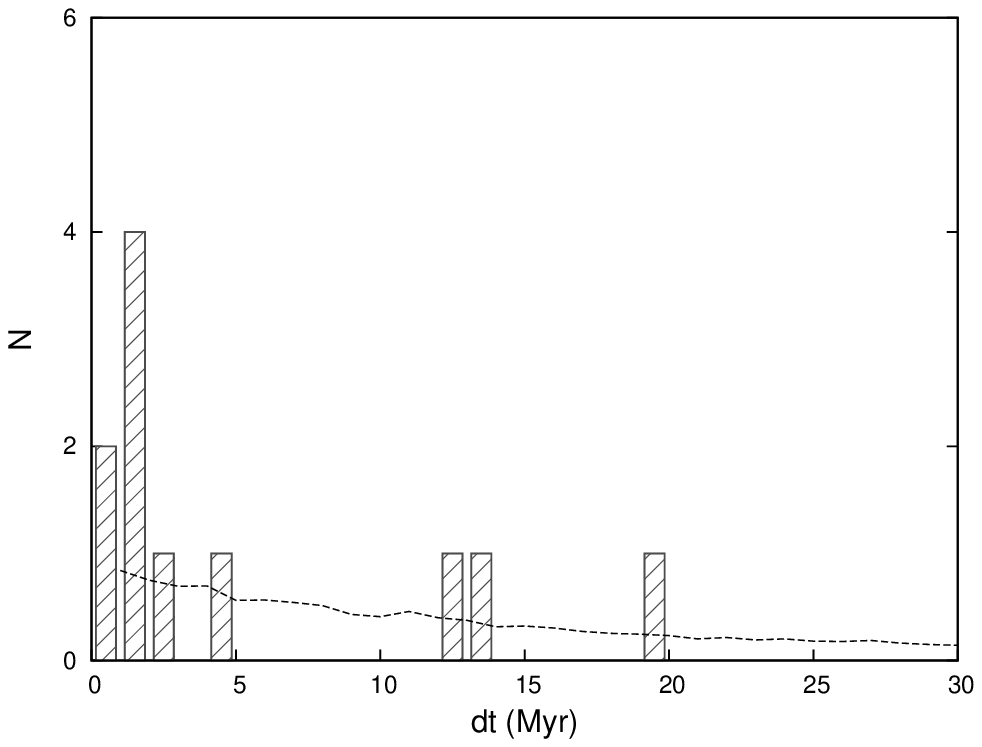}
\end{minipage}
\caption{Distribution of differential ages for the 41 impact craters  (histogram),  and for 1000 synthetic datasets extracted from the smoothed (Rice) distribution, normalised to the real dataset. Top to bottom: craters in the age ranges 0-35, 35-500 and 100-500~Myr respectively. An excess of clustering up to 2~Myr (relative to a random distribution) is found throughout the record. }
\label{fig:div}
\end{figure}

Fig.~\ref{fig:kernel} is a plot of the probability density distribution of the listed impact structures, using Gaussian kernels corresponding to the individual age errors. The Figure reveals that there is a tendency for impact craters to cluster in time.  Fig.~\ref{fig:div} shows the differential ages of the real dataset along with the average of 1000 equivalent datasets obtained by random extraction from the erosion curves (which of the two curves of Fig.~\ref{fig:craters} is employed makes no appreciable difference to the outcome).  A slight excess of crater pairs relative to a random distribution is evident, and persists throughout the timespan of the dataset.  Table~\ref{ta:craterlist} reveals that 26 of the 41 impact structures listed belong to about nine temporal clusters with $\delta t\la 2$~Myr, each of whose members are generally of identical age to within their standard deviations.  Rejecting impact craters ages over 4~Myr from the mid-point of the nearest cluster yielded the weighted mean cluster ages listed in Table~1.  The maximum deviation from the weighted mean within each episode averages 1.6~Myr.

\begin{figure}
\fontsize{12}{12} \selectfont
\includegraphics[angle=0,width=\linewidth]{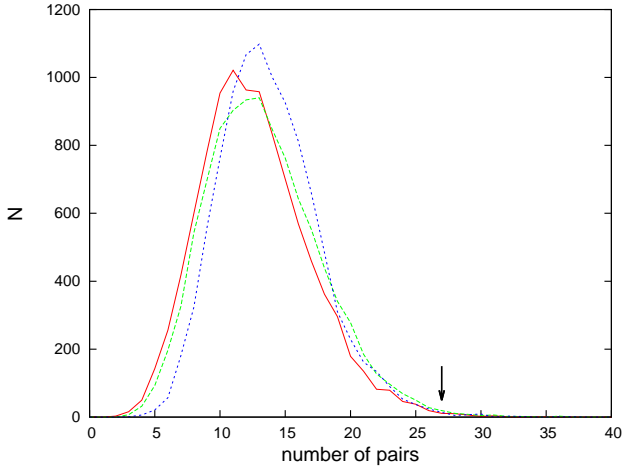}
\caption{Three simulations, each involving 10000 trials, in which pairs of craters are counted having age difference $\delta t \leq$2~Myr. From highest to lowest, the curves correspond to models in which (a) the actual crater ages are randomised by up to $\pm$50~Myr; (b) erosion is fitted to bins 100~Myr wide; and  (c) fitted to bins 70~Myr wide. The actual dataset (arrow) has 27 such pairs. The synthetic datasets all yield similar probabilities $\sim 2\times 10^{-3}$ that this number would be obtained by chance.}
\normalsize
\label{fig:all}
\end{figure}

There are 27 pairs of craters with age separations $\delta t\leq2$~Myr in the dataset (a triplet of craters bunched within 2~Myr would yield three pairs).  To test whether this pair count is compatible with the random arrival of single impactors, comparison was made with expectations from synthetic data. Two models were based on  10,000 extractions in batches of 41 from each of the secularly declining curves of Fig.~\ref{fig:craters}, the number of crater pairs being counted in each dataset so constructed. A further approach was to apply random perturbations to the crater ages, enough to wipe out any clustering but not enough to significantly change the overall distribution. Ages were varied randomly by up to $\pm 50$~Myr subject to keeping the data within (0, 500)~Myr. The results are shown in Fig.~\ref{fig:all} . The number of trials where $\geq$27 pairs out of 10,000 were obtained were, for the Rice and Sturges secular fits, and for the randomized datasets, respectively (37, 57, 42). Thus broad secular trends make little difference to the counting of closely separated crater pairs, all three yielding departures from randomness with probability $\sim$99.8\%.

\begin{figure}
\includegraphics[angle=0,width=\linewidth]{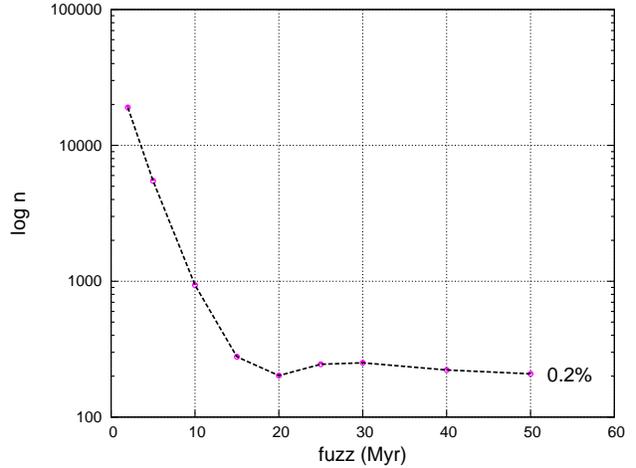}
\caption{A series of runs, in each of which 100,000 synthetic datasets are created from the real one by applying random perturbations up to a maximum (`fuzz') to individual crater ages. For each 100,000-trial run, the number $n$ of trials yielding at least 27 pairs ($\delta t <$2~Myr) is recorded. As the time resolution of the record is degraded, it asymptotically approaches a random configuration which reproduces the observed number of crater pairs in only 0.2\% of trials.}
\label{fig:fuzz}
\end{figure}

In creating synthetic datasets by applying random perturbations ${\leq} f$~Myr to the real one, any significant difference which emerges can only be due to structures on timescales ${\le}f$, since by construction the datasets are otherwise identical. In the limit $f\rightarrow 0$, synthetic and actual datasets become identical and no signal can be detected. By progressively degrading the resolution (increasing $f$), memory of the structure is progressively lost from the synthetic datasets, and an upper limit can be placed on the timescale of whatever non-random structure is present. Fig.~\ref{fig:fuzz} shows the outcome of this exercise for sets of 100,000 trials in each of which the crater ages were randomly varied by up to $f$~Myr and the number of pairs ($\delta t <$2~Myr) was counted.  Randomisations of $\pm$20~Myr or more in the crater ages yield the observed number of pairs with probability 0.2\%. However, it can be seen that even on timescales  $\la$10~Myr, nonrandom structure is present with probability $\sim$99\%.  For the 16 craters older than 100~Myr listed in Table~\ref{ta:craterlist}, the mean separation is $\sim$25~Myr and yet there are 6 crater pairs with $\delta t \leq$1.6~Myr. Randomization trials with these data yielded 6 or more such pair counts in $\sim$0.5\% of 10,000 cases.

\begin{figure}
\includegraphics[angle=0,width=\linewidth]{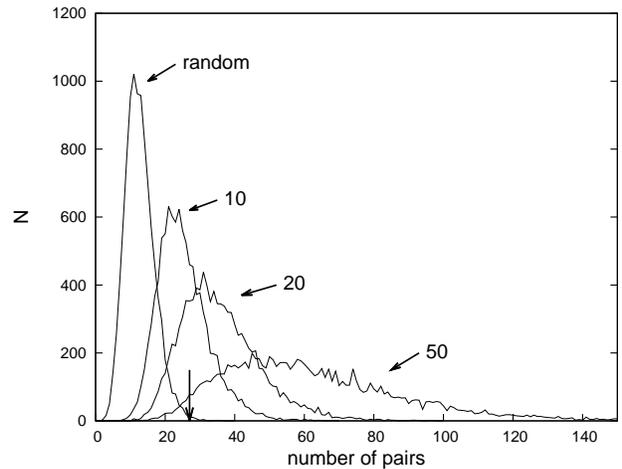}
\caption{Four simulations, each involving 10000 trials, in which pairs of craters are counted having age difference $\delta t \leq$2~Myr. The actual dataset has 27 such pairs. The curve on the left shows the distribution of such pairs when clustering is purely random. The others correspond to bombardment episodes of maximum amplitudes 10, 20 and 50, structured as described in the text.}
\label{fig:trends}
\end{figure}

Comparison was then made with models comprised of bombardment episodes, generated under the following assumptions: \\
(a) Large comets were assumed to arrive into Earth-crossing orbits (JFC or Encke) at random, and disintegrate. \\
(b) Impact craters were randomly generated from the fragments in numbers proportional to the comet mass, which was extracted randomly from a population index $\alpha\sim 1.9$ appropriate to the JFC population, to some maximum. This latter arises from a combination of physical and dynamical evolution (Di Sisto et al 2009).  For models in which impactors are the disintegration products of cascading fragmentation, $\alpha$ is probably somewhat smaller (Sekanina \& Chodas 2007; Ishiguro et al. 2009). \\
(c) The measured duration of a bombardment episode was taken to be 2~Myr, determined largely by the dating uncertainties (Table~\ref{ta:craterlist}).  \\
(d) Typically, in each trial, 200 craters were generated in this way over a 500~Myr period. They were then reduced in numbers randomly, in proportions matching the observed decline of detected craters derived from the Rice cubic fit, until 41 remained.  

Fig.~\ref{fig:trends} shows the outcome of trials in each of which 10,000 fictitious datasets were constructed. Each trial had, successively, a maximum possible amplitude $n_c =$ 2, 3, 4 \ldots for episodes of bombardment. It is clear from the Figure that the observed count of 27 pairs with $\delta t \leq 2$~Myr is rare under the hypothesis $H_0$ of no clustering, but is comfortably encompassed by a wide range of bombardment amplitudes. The probability of no impact clustering, given the dataset $D$, can be expressed as
\begin{equation}
P(H_0|D) = \frac{P(D|H_0) P_{prior}(H_0)}{P(D|H_0) P_{prior}(H_0) +  P(D|H_1)P_{prior}(H_1)}
\end{equation}
where 
\begin{equation}
P(D|H_1)P_{prior}(H_1) = \Sigma{P(D|H^\prime)P_{prior}(H^\prime)}
\end{equation}
$H_1$ represents the clustering hypothesis,  $P_{prior}(H_0)$ and $P_{prior}(H^\prime)$ represent prior probabilities in Bayesian terminology  and the summation is taken over all viable cluster models $H^\prime$ corresponding to $n_c$ = 2, 3, 4 \ldots. 

We adopt the prior hypothesis that the data are equally likely to be clustered or non-clustered. The relative probabilities are calculated from the occupancy of the $n = 27$ bin under the various hypotheses. On the no-clustering hypothesis, the bin with 27 hits has occupancy 15$\pm5$ `hits' out of 10,000 trials: $P(D|H_0) = 1.5\times 10^{-3}$. The summation of the bin 27 populations for $n_c = $ 2, 3, \ldots 100 yields  a total occupancy of $\sim$15,000 for a million trials and hence $P(D|H_1) =  1.5 \times 10^{-2} $.  Thus the posterior odds ratio $P(H_1|D)/P(H_0|D){\sim}10$, a positive result favouring the existence of clustered impacts.

\subsection{Bombardment intensities and causes}
Although the existence of impact episodes is thereby indicated, their intensity is not well constrained.  The widths of the bombardment episodes in the trials, typically 2~Myr, may reflect dating uncertainties rather than their true duration.  An amplitude of factor three above  background over 2~Myr translates to a factor of 300 over background for a comet disintegration of 20,000 years' duration, or 3000 over 2000 years. The only impact episode for which we have precise relative dating is that of 65~Myr BP. The Chicxulub  structure of that era is the largest impact crater known to have been formed over the Phanerozoic, and is generally associated with the end-Cretaceous extinctions with estimated species loss of $\sim$57--83\%.  Although originally ascribed to an asteroid impact, iridium and energy constraints yield a best estimate that the impactor was a comet 10-80 km in diameter (Durand-Manterola \& Codero-Tercero 2014).  However a detailed stratigraphic study of the Ukrainian Boltysh crater by Jolley et al (2010) has shown that it was formed only 2000-2500 years before the Chicxulub impact in the Gulf of Mexico (the radiometric age given in Table\ref{ta:craterlist} has standard  deviation $\pm$0.6~Myr). This short timescale is consistent with the breakup of a large short-period comet yielding $N$ Earth-crossing fragments with dynamical or physical lifetimes $L$~yr such that $NL{\ga}C$, where $C$ represents the mean lifetime of a fragment before collision with the Earth. For a comet in an Encke-like orbit, $C{\sim}200$~Myr and 10$^5$ km-sized fragments lasting 2000~yr (with total mass corresponding to a nominal 70 km diameter comet)  or 10$^4$ fragments lasting 20,000~yr (a 30 km comet) would suffice.  

Other contenders for bombardment episodes are comet showers from the Oort cloud, or impacts from the disintegration of asteroids in the main belt, but neither mechanism seems capable of reproducing the sharpness and amplitude of the observed episodes. Kaib \& Quinn (2009) constrain the strength of comet showers through stellar penetrations of the inner Oort cloud, by noting that too high a comet number density in the inner Oort cloud would overproduce long period comets. They find that there has been no more than about one shower of Eocene-Oligocene strength over the past 500~Myr. Terrestrial impacts from the breakup of main belt asteroids are too infrequent by order of magnitude to account for the episodes (Napier 2006). The main asteroid families generally yield 0-4  terrestrial impactors $>$1~km across, arriving over 5-90~Myr  (Zappala 1998).

\begin{table}
\centering
\begin{tabular}{rrr}  \hline
diam (km)& in    &   out  \\ \hline
$>20$      &  15   &   7    \\
10-20       &   4    &   6    \\ 
 3-10        &   6    &   3    \\ \hline       
\end{tabular}
\caption{Membership of bombardment episodes listed in Table~\ref{ta:craterlist}. There is no evidence of a systematic trend with crater diameter.}
\label{tab:inout}
\end{table}

The question arises whether the tendency to clustering is a function of crater diameter. In Table~\ref{tab:inout} the craters are partitioned by size and by membership or otherwise of an impact cluster. No significant correlation is evident between crater size and cluster membership: the tendency for impactors to  arrive in groups extends down to sub-kilometre bodies.

\subsection{The most recent episode}
The apparent spike in the bombardment rate of the past few million years could be seen as due simply to the relative completeness of discovery of recent craters. However, between the eight well-dated craters $\le 5$~Myr in age and the Chesapeake/Popigai group at $\sim$36~Myr,  there is only one large impact structure, the 24~km Nordlinger Ries at 15~Myr (from an achondritic impactor). The Earth Impact Database, with 184 craters, yields only one additional crater, the 3.6 km diameter Colonia with an age in the range 5$<t<$35~Myr.  On a random impact scenario, the lunar cratering record indicates that bodies large enough to make 20 km craters strike the Earth every 600,000 years (Bland 2005). Allowing for ocean impacts, there is then an expectation that $\sim$17 land craters of at least this size should have been formed over the 30~Myr interval (Bland 2005), most of which should have been discovered since this a geologically young period. The gap therefore appears to be real.

\section{Comparison with marine extinction rates}

\begin{figure}
\includegraphics[angle=270,width=\linewidth]{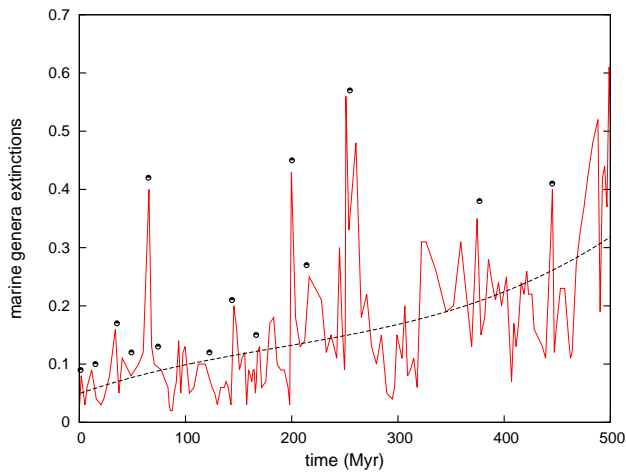}
\caption{Bombardment episodes (asterisks) and relative extinction intensities of marine genera from Bambach (2006). Updated from Napier (2014).} 
\label{fig:corr}
\end{figure}

\begin{table}
\centering
\begin{tabular}{rrrll} \hline
No. &  extinction &  episode &  $\Delta t$  & boundary \\ \hline
1    &    1.8         &   1.1      &   0.7   &     \\
2    &  33.9         &  35.4     &   1.5   &  end Eocene   \\
3    &  65.5         &   65.0	  &   0.5   &   end Cretaceous \\
4    &  93.5         &   --     	  &   4.2   &    \\
5    & 145.5        &  143.7	  &   1.8   &  end Jurassic   \\
6    & 183.0        &    --        &    --    &    \\
7    & 199.6        &   200.2	&   0.6  &   end Triassic  \\
8    & 251.0        &   254.7   &  3.7   &    end Permian \\
9    & 260.4        &   --        &   --    &    \\
10  & 322.2        &    --	     &   --    &    \\
11  & 359.2        &    --       &   --    &   \\
12  & 374.5        &  377.2	 &   2.7  &   end Devonian\\
13  & 385.3        &   --	       &   --    &    \\
14  & 443.7        &   445.2 &   1.5  &    \\
15  & 445.5        &   445.2 &   0.3  &    end Ordovician \\
16  & 488.3        &   --	      &   --     &   \\
17  & 494.7        &   --	      &   --     &   \\
18  & 498.9  	    &   --	      &   --         \\ \hline				\end{tabular}
\caption{Epochs (Myr ago) of marine genera extinction events considered by Melott \& Bambach (2010) to be genuine mass extinctions, along with those impact episodes of Fig.~\ref{fig:kernel} occurring within $\Delta t <$5~Myr of the extinctions. Seven of the twelve such events occur at major geological boundaries.}
\label{ta:compare}
\end{table}

For comparison with extinction data, the 500~Myr marine fossil record database assembled by Bambach (2006) was employed. This came from the Paleobiology Database and the compendium by Jack Sepkoski: discussion and references are found in Melott \& Bambach (2010). The impact crater ages employed have been determined radiometrically, whereas the marine extinctions are generally based on stratigraphy and contain rounding errors due to the assignment of extinctions to the nearest geological substage. The relative extinction intensities of genera are plotted in Fig.~\ref{fig:corr}, along with the bombardment episodes. According to Bambach, 18 events in the last 500~Myr may properly be described as mass extinctions. In Table~\ref{ta:compare} these extinction peaks are listed along with those bombardments which occur within $\pm$5~Myr of a peak. The well-dated single impacts in Fig.~\ref{fig:kernel} are included. Two of the mass extinctions, events 14 and 15, are extremely close together and share a bombardment episode. For statistical purposes they are here counted as one. If we regard each bombardment episode as a single event and include the single impacts, there are 14 impact episodes, seven of which lie within $\pm$3~Myr of an extinction episode, six within $\pm$2~Myr and four within $\pm$1~Myr. Seven coincide with the terminations of major geological boundaries to within the dating errors, namely the Ordovician, Devonian, Permian, Triassic, Jurassic, Cretaceous and Eocene. The bombardments comprise 11 multiple episodes (Table~\ref{ta:craterlist}) and three singles.  

Monte Carlo trials were carried out in which 14 synthetic bombardment episodes were randomly extracted over 500 million years, and the number which lay within $\Delta t$~Myr  of the Bambach mass extinction (Table~\ref{ta:compare}) was recorded. Not more than one episode could be assigned to an extinction, since they would not be distinguishable.  In reality, uncertainty attaches also to the precise dating of the extinctions, and this may be soaked up in $\Delta t$. Trials for each of the thresholds $\Delta t = \pm$(3,2,1)~Myr yielded probabilities (0.032, 0.005, 0.012) that the coincidences are due to chance. Excluding the first 30~Myr  of the extinction and bombardment records (in case the recent `episode'  is actually discovery bias), makes no essential difference to the probabilities. These coincidences independently validate the reality of the cratering episodes since, if they were statistical artefacts, there would be no reason to expect them to correlate with the observed mass extinctions. They also demonstrate that bombardment episodes have been major factors in the evolution of life since the pre-Cambrian explosion.

There is continuing debate amongst palaeontologists about whether the Cretaceous-Palaeogene extinctions were sudden and caused by the Chicxulub impact, or whether there was a more continuous decline over $\la$1~Myr of species caused by global cooling, sealevel fluctuations and the main phase of the Deccan vulcanisms (MacLeod 2013). No other mass extinction has yet been identified with a single large crater, and most or all of the impact peaks are due to impactors too small to have caused major mass extinctions on their own. For example, the Araguainha crater in central Brazil is only 40 km across and was itself incapable of causing the near-extinction of life at the Permo-Triassic boundary 250 million years ago; the end-Ordovician peak at 445~Myr represents two craters only a few kilometres in diameter; the Triassic peak at 200~Myr represents two impact structures with diameters in the  probable range 20-40 km, and so on. The collectively strong association of such minor impactors with major extinctions implies that they are proxies rather than individual causes of the mass extinctions.

\section{Terrestrial consequences}
\label{sec:terr}

The stepwise nature of several mass extinctions has often been noted, and the hypothesis of multiple impacts as their cause, based on purely biostratigraphic considerations, is not new (e.g. the Late Devonian crisis of 365-368 Myr BP: McGhee 1996). Some of these ancient events were highly destructive, with intensities that came close to extinguishing life on Earth. There is, however, a spectrum of intensities, as there is a spectrum of comet masses, and the correlation evident in Fig.~\ref{fig:corr} and Table~\ref{ta:compare} suggests that episodes of bombardment may exert a more continuous control over the evolution of species than would be expected simply from rare, isolated impacts. A progressive `toughening' of marine genera with time is evident from the slope of the curve in Fig.~\ref{fig:corr}. There is no obvious astrophysical explanation for this decline in extinction rate, but no need for one as several palaeontological hypotheses have been advanced to account for it (MacLeod 2013). 

Birks et al (2007) have pointed out that the most damage which can be done to the biosphere with the least impact energy is to deplete the stratospheric ozone layer, so increasing the level of UVB radiation reaching the surface of the Earth. Depletion by way of impact occurs partly through the activation of halogens thrown into the stratosphere through salt-water impact, and partly through the injection of NO created in the impact fireball. They consider that the impact of a $\sim$0.5~km asteroid would have serious implications for the biosphere, and that such impacts occur at $\sim$0.1~Myr intervals. The effects of enhanced UVB on terrestrial organisms have been discussed by numerous authors. Melott \& Thomas (2011) have pointed out that a current anthropogenic 3\% depletion in global stratospheric ozone is causing measurable damage to the biosphere (H\"{a}der et al 2007). They consider that a 30\% depletion is well above the lethality level for phytoplankton, which lie at the base of the oceanic food chain. Such a depletion would cause a major marine extinction event. Melott et al (2010) have also pointed out that cometary airbursts may create nitrous oxides in the stratosphere, depleting ozone. In the present scenario, bombardment episodes may enhance the frequency of small impacts by two or three powers of 10 over background during the disintegration of a giant short-period comet.

A further possible effect is climatic perturbation due to the dust production expected in the course of disintegration. If cometary fireballs are reduced to micron-sized particles on entering the atmosphere (Klekociuk et al 2005), then a current infall of cosmic dust of 40,000$\pm$20,000 tons a year (Love \& Brownlee 1993) yields an atmospheric optical depth $\tau{\sim}$8$\times 10^{-3}$ for an absorption coefficient $\kappa{\sim}10^{-5}$\,cm$^2$\,g$^{-1}$. This figure may rise to $\tau{\sim}$0.2 over the course of a giant comet disintegration (Napier 2001), raising the prospect of repeated pulses of global cooling each of millennia in duration.

\section{Summary and conclusions}

As with the periodicity issue, the analysis of bombardment episodes is inhibited by small number statistics. Nevertheless positive evidence for such episodes appears in the impact cratering record. They are found to be tightly correlated with substantial marine extinctions at the level of genera. To date only the Cretaceous-Tertiary extinctions have been widely identified as being caused by a single large impact, and even that claim is disputed by many palaeontologists (Keller et al 2012, MacLeod 2013, Keller \& Kerr 2014). Several large impact craters, which might have been expected to have caused mass extinctions, are associated with only minor, regional ones. The main asteroid belt seems inadequate to supply the larger impactors, and large cometary populations have now been revealed extending to beyond the planetary system, with the potential to be thrown into the near-Earth environment. The top-heavy mass distribution of these comets, their likely disintegration history when thrown into short-period orbits, and the potential of their debris to cause deleterious effects, make them likely candidates for the mass extinctions observed in the terrestrial record and the associated geological disturbances. The mechanisms involved are a matter for detailed stratigraphic studies; it is suggested here that prolonged atmospheric perturbations arising from fireball storms and dusting are the most energy-efficient means of collapsing food chains, yielding both marine and land extinctions.

\section{Acknowledgements}

I am indebted to Duncan Steel, and the referee Coryn Bailer-Jones, for thorough critiques of earlier drafts of this paper. Richard Bambach kindly supplied the palaeontological data, and the impact crater data compendium came from the Earth Impact Database created and maintained by the staff of the Planetary and Space Science Centre at the University of New Brunswick, Canada.

\end{document}